\documentclass[useAMS,usenatbib,usegraphicx]{mn2e}

\newcommand{\kev}{keV}

\newcommand{\fe}{Fe~K$\alpha$}
\newcommand{\etal}{et al.}
\newcommand{\fouru}{4U~1820--30}

\title[Reflection from a disc during a neutron star burst]
  {Reflection spectra from an accretion disc illuminated by a
  neutron star X-ray burst}
\author[D.\ R.\ Ballantyne]
  {D.~R.~Ballantyne\thanks{ballantyne@cita.utoronto.ca}\\
  Canadian Institute for Theoretical Astrophysics, 60 St. George
  Street, Toronto, Ontario, Canada M5S 3H8}

\pagerange{\pageref{firstpage}--\pageref{lastpage}}
\pubyear{2004}

\usepackage{times}

\begin{document}

\label{firstpage}

\maketitle

\begin{abstract}
Recent time-resolved X-ray spectra of a neutron star undergoing a
superburst revealed an \fe\ line and edge consistent with reprocessing
from the surrounding accretion disc. Here, we present models of X-ray
reflection from a constant density slab illuminated by a blackbody,
the spectrum emitted by a neutron star burst. The
calculations predict a prominent \fe\ line and a rich soft X-ray line
spectrum which is superimposed on a strong free-free continuum. The
lines slowly vanish as the ionization parameter of the slab is
increased, but the free-free continuum remains dominant at energies
less than 1~\kev. The reflection spectrum has a quasi-blackbody shape
only at energies greater than 3~\kev.  If the incident blackbody is
added to the reflection spectrum, the \fe\ equivalent width varies
between 100 and 300~eV depending on the ionization parameter and the
temperature, $kT$, of the blackbody. The equivalent width is
correlated with $kT$, and therefore we predict a strong \fe\ line when
an X-ray burst is at its brightest (if iron is not too ionized or the
reflection amplitude too small). Extending the study of reflection
features in the spectra of superbursts to lower energies would provide
further constraints on the accretion flow. If the \fe\ line or other
features are relativistically broadened then they can determine the
system inclination angle (which leads to the neutron star mass), and,
if the mass is known, a lower-limit to the mass/radius ratio of the
star.
\end{abstract}

\begin{keywords}
accretion, accretion discs --- line: profiles --- radiative transfer
--- X-rays: binaries --- X-rays: bursts
\end{keywords}

\section{Introduction}
\label{sect:intro}
The reprocessing of external X-rays by accreting material has been
observed from active galactic nuclei (AGN) and Galactic black hole
candidates (GBHCs) for over a decade
\citep*[e.g.,][]{pou90,np94,gie99,bvf03,mil03}. In these objects, the
origin and location of the illuminating X-ray power-law is largely
unknown, although it is widely believed to be either within a magnetic
accretion disc corona \citep*[e.g.,][]{gal79,haa93,haa94,dm98} or a
centrally located geometrically thick flow
\citep*[e.g.,][]{sle76,zdz99,zlgr03}. Nevertheless, many models of
X-ray reflection from black hole accretion discs have been computed
\citep*{gf91,mpp91,ros93,zyc94,ros99,nkk00,nk01,brf01,roz02,btb04},
and, in some cases, successfully applied to data
\citep*{bif01,orr01,der02,lon03,bdn03,mil03}.

Recently, \citet{sb02} discovered an \fe\ line and edge in the X-ray
spectra of the low-mass X-ray binary (LMXB) \fouru\ as it was
undergoing a superburst --- a powerful thermonuclear explosion on the
surface of the neutron star that can last many hours
\citep*{corn00,sb03,kuul03}. Based on the strength of the \fe\ line,
these authors suggested that the features were caused by reflection of
the burst blackbody spectrum from the surrounding accretion
disc \citep[e.g.,][]{dd91}. This hypothesis was supported by spectral
fitting with detailed reflection models \citep{bs04}. The unique
aspect of reflection during a burst from a LMXB is that there is no
uncertainty in the location of the X-ray source, as it is located on
the surface of the neutron star and outshines the persistent X-ray
emission from the disc. This fact allows for far less ambiguity when
analyzing changes to the reflection features, as they are more likely
to trace the evolution in the accretion flow \citep{bs04}.

X-ray emission lines have also been observed in the persistent
emission from LMXBs \citep[e.g.,][]{sma93,ang95,sch99,adnm00}, and
these are also thought to be caused by reprocessing by matter in the
accretion flow \citep{ss73,mmh82}. The spectral shape of the
persistent emission is consistent with bremsstrahlung with a
temperature of 5--10~\kev\ \citep*{lph01}, most likely arising from the
boundary layer between the disc and the neutron star. Detailed models
of the photoionized layer on the disc have been performed and predict
a wealth of emission lines in the soft X-ray band
\citep*[e.g.,][]{kk94,jrl02} that can be compared against observations
made with the grating spectrometers onboard \textit{Chandra} and
\textit{XMM-Newton} \citep[e.g.,][]{sch01,cott01,kabc03}. In
contrast, a X-ray burst emits a blackbody spectrum
\citep*{swa77,hld77}, and can reach the Eddington luminosity
($L_{\mathrm{Edd}} \sim 10^{38}$~erg~s$^{-1}$), well over an order of
magnitude greater than the persistent luminosity of many LMXBs
\citep{sb03}. Thus, disc reflection will be dominated by the blackbody
component during the burst.

This paper presents reflection spectra from a uniform accretion disc
illuminated by a blackbody, as in a LMXB during a Type~I X-ray
(super)burst. These spectra were used to fit the superburst data of
\fouru\ \citep{bs04}, but, as shown below, they contain much more
information than what was actually needed for those \textit{Rossi
X-ray Timing Explorer} (\textit{RXTE}) data. Observations of
reflection features from Type~I bursts or superbursts with sensitive,
large bandwidth instruments such as the X-ray Telescope (XRT) on
\textit{Swift} could provide a wealth of new information on the
structure and behaviour of accretion discs.

In the next section we outline the reflection calculations, and
present the resulting spectra in Section~\ref{sect:res}. We conclude
by discussing the results in Section~\ref{sect:discuss}.

\section{Calculations}
\label{sect:calc}
The reflection spectra were computed using the code of
\citet{ros93}. Details of the computational techniques can be found in
that paper and in references therein (see also \citet{ros99} and
\citealt{brf01}), and so only a brief overview will be presented here.

A one-dimensional slab of material with a constant hydrogen number
density $n_{\mathrm{H}}$ is illuminated by a blackbody spectrum with
temperature $kT$ and flux (defined between 1~eV and 100~\kev)
$F_{\mathrm{X}}$. Thus, we define the ionization parameter
\begin{equation}
\xi = {4 \pi F_{\mathrm{X}} \over n_{\mathrm{H}}}.
\label{eq:xi}
\end{equation}
The input radiation is transferred into the layer using the two-stream
approximation \citep{ryl79} while the outgoing radiation is treated using the
modified Fokker-Planck/diffusion method of \citet*{rwm78}. Therefore,
the output spectrum is averaged over all viewing angles. The ionization
and thermal structure of the illuminated gas is iterated until each
zone reaches ionization and thermal balance, and at this point the
final reflection spectrum is computed. The illuminated slab always has
a large enough Thomson depth ($\tau_{\mathrm{T}} =$6--20, depending on
$\xi$) so that the highest energy photons will interact many times with the
gas. The cosmic abundances of \citet{mcm83} were assumed, but hydrogen
and helium are taken to be completely ionized at all times (i.e.,
$n_{e}=1.2n_{\mathrm{H}}$). The following ionization stages of the
most important astrophysical metals are treated in the calculation:
C~\textsc{v}--\textsc{vii}, N~\textsc{vi}--\textsc{viii},
O~\textsc{v}--\textsc{ix}, Mg~\textsc{ix}--\textsc{xiii},
Si~\textsc{xi}--\textsc{xv} and Fe~\textsc{xvi}--\textsc{xxvii}.

Models were calculated for $1 \leq kT \leq 3.15$~\kev, as this spans
the range most often observed during X-ray bursts \citep{kuul03}, and
for $1 \leq \log \xi \leq 3.75$. To change $\xi$, $n_{\mathrm{H}}$
was fixed at 10$^{15}$~cm$^{-3}$, and $F_{\mathrm{X}}$ was varied.

\section{Results}
\label{sect:res}
\begin{figure*}
\begin{minipage}{180mm}
\centerline{
\includegraphics[width=\textwidth]{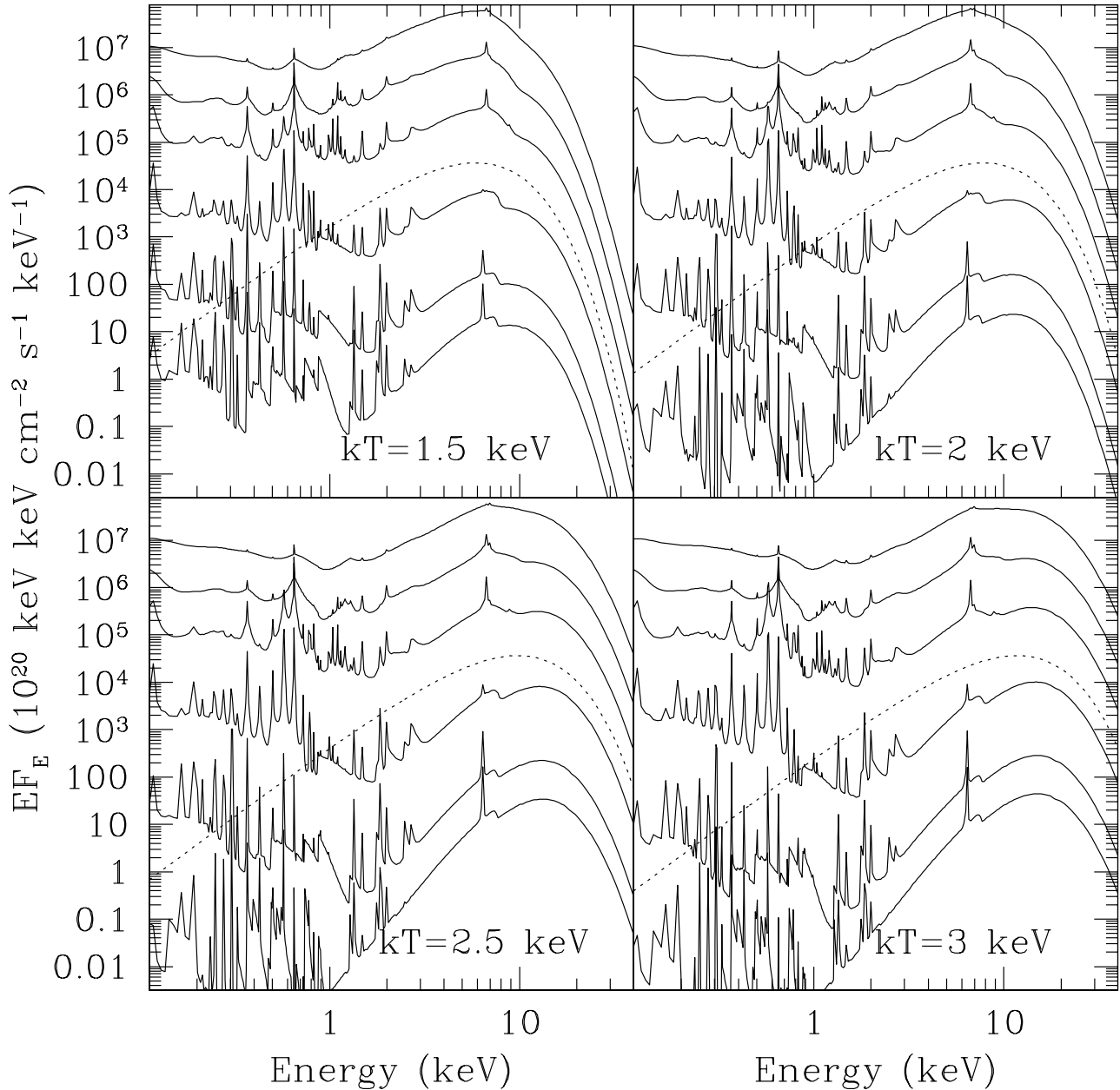}
}
\caption{Examples of reflection spectra from an accretion disc
  illuminated by a blackbody continuum. The four panels show results
  for different values of $kT$. Within each panel 6 reflection spectra
  are shown with solid lines. They denote models with (moving from
  bottom to top) $\log \xi =$ 1, 1.5, 2, 2.5, 3, 3.5. All of the
  spectra, except for the $\log \xi =2$ model, have been vertically
  offset from one another for clarity. The dotted line in each panel
  shows the illuminating blackbody for the $\log \xi=2$ model.}
\label{fig:spectra}
\end{minipage}
\end{figure*}

We begin by considering the evolution of the reflection spectra as a
function of $\xi$ (cf. Fig. 2 in
\citealt{ros99}). Figure~\ref{fig:spectra} shows a series of spectra
for four different values of $kT$.
Each panel shows reflection spectra with $\log \xi =$ 1, 1.5, 2, 2.5,
3, 3.5 (from bottom to top), thereby spanning the range from very
weakly ionized reflectors to highly ionized ones. For the $\log \xi = 2$
model, we have also plotted the incident blackbody spectrum. 

Unlike the power-law continua of AGN and GBHCs where the radiant
energy per frequency decade reaches a maximum at energies $> 10$~\kev\
(for photon-indices $\Gamma < 2$) where atomic absorption is minimal,
the blackbody spectra peak at energies where the photoelectric
absorption by metals is very important. As a result, only a small
fraction of the photons incident on the slab are scattered back
without first being absorbed by an ion. The reflection spectra
therefore only mimics a blackbody at energies $E \ga 3$~\kev, with the
agreement increasing as $\log \xi$ grows and the metals are
ionized. The energy absorbed by the metals in the gas is thermalized
and reemerges as a strong soft free-free continuum with a variety of
recombination lines superimposed. As this thermal emission includes the
power absorbed at higher energies, it outshines the incident blackbody
at $E \la 1$~\kev\ for all values of $\xi$ considered. The rich
emission line spectrum is reminiscent of those predicted from models
of reprocessing of the persistent emission \citep{jrl02}. However, in the most
luminous stage of an X-ray burst, the disc may be severely ionized
\citep{bs04}, and will leave a Compton broadened O~\textsc{viii}
Ly$\alpha$ line as the only emission feature in the soft X-ray region
of the spectrum.

\begin{figure}
\centerline{
\includegraphics[width=0.5\textwidth]{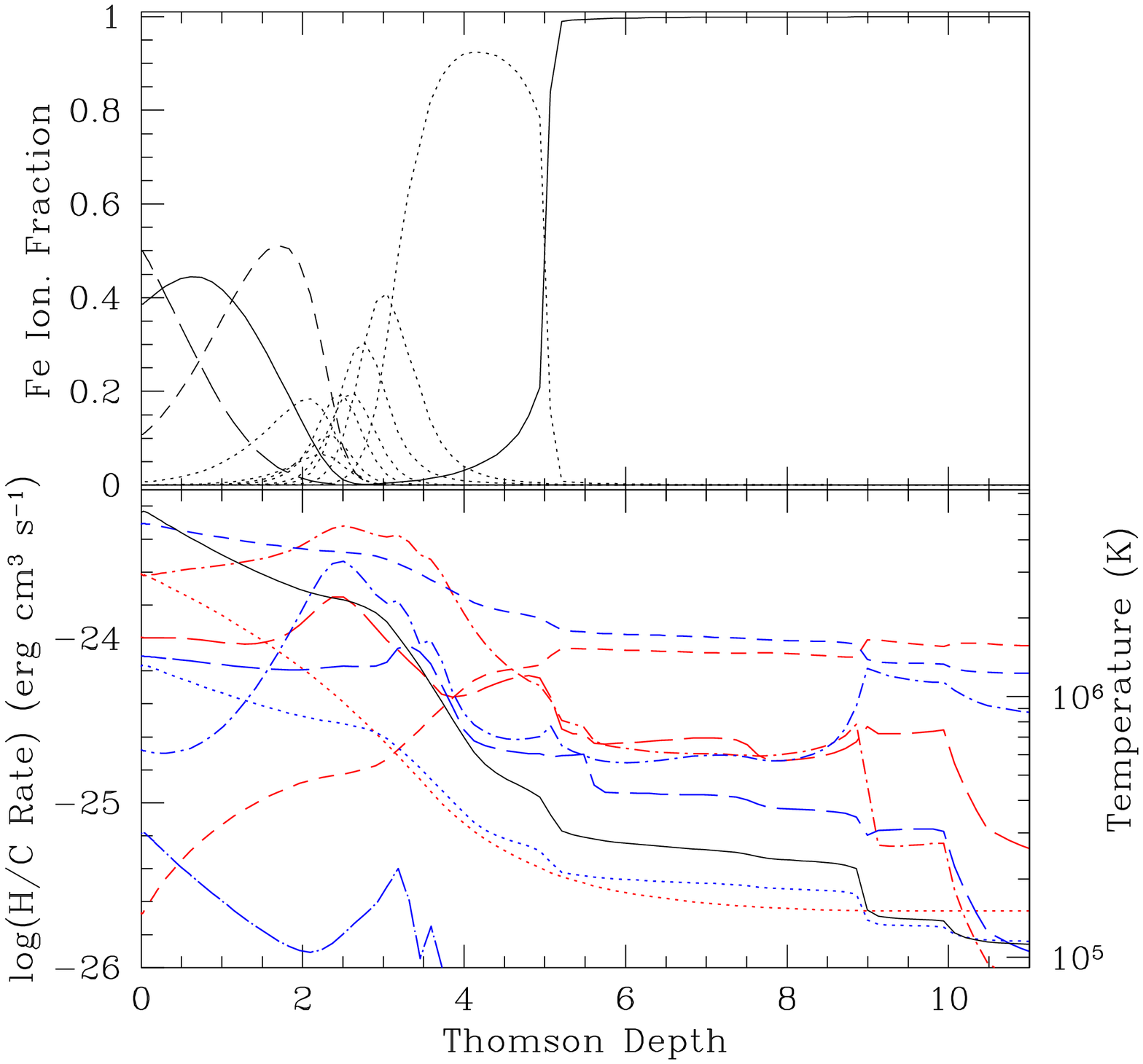}
}
\caption{Details from the $\log \xi =3$, $kT=2.5$~\kev\ reflection
  calculation as a function of Thomson depth into a slab with
  $n_{\mathrm{H}}=10^{15}$~cm$^{-3}$. The upper panel plots the iron
  ionization structure. Moving from the surface down into the slab,
  the iron ions encountered are Fe~\textsc{xxvii} (long-dashed line),
  hydrogenic Fe~\textsc{xxvi} (solid line), Fe~\textsc{xxv}
  (short-dashed line), Fe~\textsc{xvii}--\textsc{xxiv} (dotted lines)
  and Fe~\textsc{xvi} (solid line; the most neutral iron ion
  treated). The lower panel plots the (H)eating (red curves) and
  (C)ooling (blue curves) rates and the gas temperature (right-hand
  axis; black solid line). The line types distinguish the various
  physical processes: Compton heating/cooling (dotted lines),
  free-free heating/cooling (short-dashed lines), recombination
  heating/cooling (long-dashed lines), photoelectric heating/line
  cooling (dot-short-dashed lines), and three-body interactions
  (dot-long-dashed lines).}
\label{fig:rates}
\end{figure}

It is instructive to look at the most important heating and cooling
processes within the illuminated layer. A plot of the heating and
cooling rates as a function of Thomson depth into the slab for the
$\log \xi = 3$, $kT=2.5$~\kev\ model is shown in the lower panel of
Figure~\ref{fig:rates}.
In this figure, the heating and cooling processes are shown by red and
blue lines respectively, while the different line styles separate the
physical process (for example, Compton heating is denoted by the red
dotted line, but Compton cooling is shown with the blue dotted
line). The upper panel of Fig.~\ref{fig:rates} shows the final iron
ionization structure in the slab. Hydrogenic iron dominates between
the surface and the first Thomson depth, but recombines to He-like Fe
before reaching $\tau_{\mathrm{T}}=2$. The iron recombines rapidly
just beyond this point and reaches the lowest ionization stage treated
(Fe~\textsc{xvi}) at $\tau_{\mathrm{T}} \approx 5$. Turning to the
heating and cooling rates, we see that free-free emission is the
dominant cooling mechanism over the entire depth of the slab. At the
surface, where the gas temperature reaches $T \sim 5\times 10^6$~K
(the black line in the lower panel of Fig.~\ref{fig:rates}), the
bremsstrahlung cooling is balanced by Compton and photoelectric
heating. However, the Compton heating rate falls rapidly with
$\tau_{\mathrm{T}}$ because there are very few high energy photons in
the incident blackbody spectra. In contrast, the photoelectric heating
rate increases with $\tau_{\mathrm{T}}$ until it reaches a maximum at
$\tau_{\mathrm{T}} \sim 3$, where iron is rapidly recombining (upper
panel of Fig.~\ref{fig:rates}) and the photoionization rates
jumps. The line cooling rate also increases rapidly here, but cannot
overcome the free-free cooling. After the iron has almost fully
recombined at $\tau_{\mathrm{T}} \sim 5$, the photoelectric heating
rate falls (very few ionizing photons can penetrate to such depths)
and the dominant heating process becomes free-free absorption. The
line-cooling rate jumps one more time at $\tau_{\mathrm{T}} \sim 9$
where oxygen is recombining (note the corresponding fall in the gas
temperature). The dominance of free-free cooling and photoelectric
heating in the thermal structure of the illuminated slab emphasizes the
dominance of the absorption-thermalization process in these
models. The observational outcome is the rich X-ray line spectrum seen
below 1~\kev\ in Figure~\ref{fig:spectra}.

Similar to the reflection spectra predicted for the environment of
accreting black holes, the \fe\ line is the most important spectral
feature above $\sim 3$~\kev\ in the blackbody models, and is an
important indicator of the ionization parameter of the reflecting
medium. Fig.~\ref{fig:spectra} shows that in each panel the \fe\ line
evolves from a strong 6.4~\kev\ line to a weak, Auger-destroyed line,
and then to a broad 6.7~\kev\ feature before disappearing entirely due
to the complete ionization of iron. This is the same sequence as that
presented by \citet{ros99} for illumination by a power-law.

\begin{figure}
\centerline{
\includegraphics[width=0.5\textwidth]{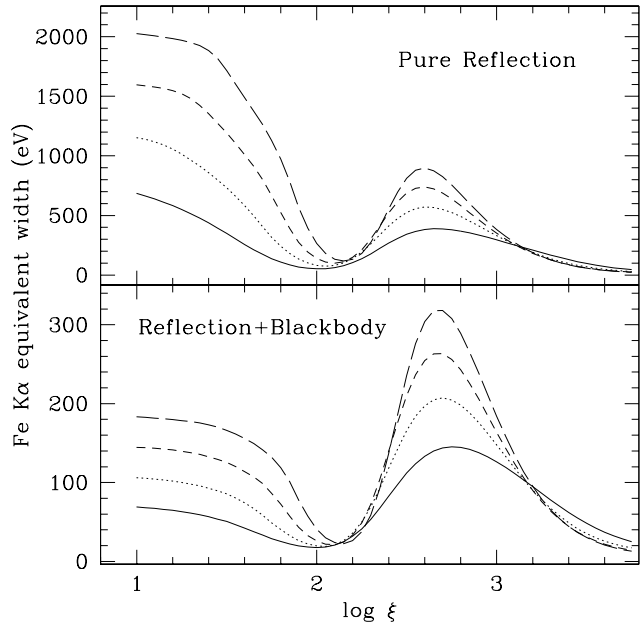}
}
\caption{The \fe\ equivalent width (EW) as a function of $\log \xi$ for the
  four values of $kT$ shown in Figure~\ref{fig:spectra}. The upper
  panel plots the EW measured from the pure reflection spectrum (i.e.,
  there is no contribution from the blackbody). The lower panel plots
  the EW measured from the sum of the reflected and incident
  spectra. In both panels the different line types separate the
  values of $kT$: 1.5~\kev\ (solid line), 2~\kev\ (dotted line),
  2.5~\kev\ (short-dashed line), and 3~\kev\ (long-dashed line).}
\label{fig:ews}
\end{figure}

The equivalent width (EW) of the \fe\ line as a function of
$\log \xi$ is shown in Figure~\ref{fig:ews}.
The top panel shows the EWs measured from the reflection spectra
themselves, while the lower panel includes the diluting effect of
adding in the incident blackbody that will also be observed along with
the reflection spectrum. The measurements were made for the four
values of $kT$ shown in Fig.~\ref{fig:spectra}, but for 75 values of
$\log \xi$ between $1$ and $3.75$. There are two peaks in
the EW evolution: one when $\log \xi \approx 1$ and \fe\ is at
6.4~\kev, and the other when $\log \xi \approx 2.7$ and the line is at
6.7~\kev. The EW of the cold line is much stronger than the ionized
one in the case of pure reflection, but is switched when the incident
blackbody is included. This is because at larger values of $\xi$ there
is less metal absorption in the illuminated slab and the reflection
spectrum is not as diluted by the blackbody as in the low $\xi$
models.

The \fe\ EW is a strong linear function of the temperature of the
blackbody; for example, when $\log \xi =1$ and the incident blackbody
is included, the EW increases from $\sim 70$~eV (at $kT=1.5$~\kev) to
$180$~eV (at $kT=3$~\kev). This is a result of the increase in photons
above the iron K edge (at $7.1$--$9$~\kev) as $kT$ is increased. This
has important implications for the study of X-ray bursts, because the
blackbody emission is at its hottest when the burst is
brightest. Therefore, unless iron is fully ionized by the burst (i.e.,
$\log \xi > 3.7$), the \fe\ line should be most prominent at the
point where the burst is most easily observable. However, a superburst
may cause changes in the accretion geometry close to the neutron
star which would reduce the reflection amplitude and the EW of the
\fe\ line \citep{bs04}.

\section{Discussion}
\label{sect:discuss}
Searches for spectral features during X-ray bursts from neutron stars
have been ongoing for many years. Since the explosion is occurring on
the surface of the star, the spectral features would be expected to be
redshifted and would therefore be a measure of the mass-to-radius
ratio of the star (via $1 + z = (1-GM/c^2 R)^{-1/2}$), a number needed
to constrain the many possible equations of state of nuclear
matter. During the 1980s there were a few reports of X-ray absorption
lines at 4.1~\kev\ during Type~I X-ray bursts
\citep*{wak84,nit88,mag89}, which were interpreted as redshifted
Ly$\alpha$ absorption from He-like iron. However, the equivalent width
of these absorption lines were on the order of hundreds of eV, which
theoretical models of spectral formation in burst atmospheres could
not reproduce \citep{frf87,dfr92}. Observations by more sensitive
instruments on \textit{RXTE} and \textit{BeppoSAX} have been unable to
discover any other examples of a 4.1~\kev\ absorption line. Recently,
the Reflection Grating Spectrometer (RGS) on \textit{XMM-Newton} has
uncovered evidence for redshifted absorption lines during the X-ray
bursts of EXO~0748-676 \citep{cpm02}, although data from 28 bursts
were needed in order to obtain the minimum signal-to-noise necessary
to find the lines.

As first suggested by \citet{dd91}, X-rays from an explosion on
a neutron star can be reprocessed by the surrounding accretion disc,
leading to features such as an \fe\ line and edge. The analysis of
these features can lead to many new insights about the structure and
evolution of the accretion disc \citep{bs04}, as well as basic
information about the neutron star itself. For example, if the
emission lines are relativistically broadened then modeling can
reveal the inclination angle of the system to the line of sight \citep{fab89},
a parameter often needed to obtain masses of the neutron star and its
binary companion. Modeling of relativistic lines also produce a
radius where the reflection originates (assuming some emissivity
profile). This radius is measured in gravitational units ($GM/c^2$),
so if the mass of the neutron star is known, the radius can be
converted to physical units, allowing a determination of a
mass-to-radius ratio. Since the reflection may occur very close to
the surface of the neutron star ($10$--$20$~$GM/c^2$ in 4U~1820-30;
\citealt{bs04}), the measurement may provide a useful lower-limit to
the $M/R$ ratio for the neutron star.

\begin{figure}
\centerline{
\includegraphics[width=0.5\textwidth]{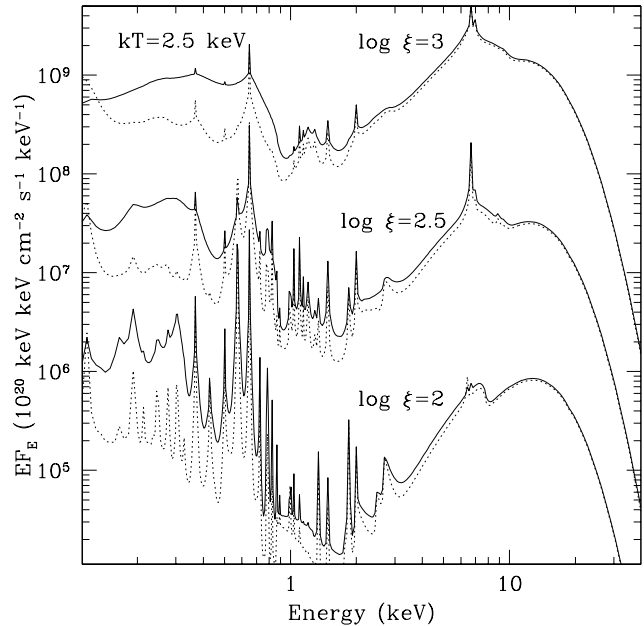}
}
\caption{The solid lines show three reflection spectra computed with a
  density of $n_{\mathrm{H}}=10^{18}$~cm$^{-3}$, while the dotted
  lines denote models with identical $\xi$ but with
  $n_{\mathrm{H}}=10^{15}$~cm$^{-3}$. A $kT=2.5$~\kev\ blackbody was
  the illuminating continuum for all cases. The increase in density
  results in only minor differences in the spectra at energies $\ga 3$~\kev,
  but there are significant changes at lower energies.}
\label{fig:densitycompare}
\end{figure}

\textit{RXTE} has so far been the most successful telescope to obtain
time-resolved spectroscopy of X-ray bursts, especially of the more
energetic superbursts \citep[e.g.,][]{sb02}. The telescope produces
data for energies greater than 3~\kev\ which allows a good
determination of the blackbody spectrum and, if detected, the \fe\
line. Yet, the reflection models predict a wealth of information at
energies less than 3~\kev\ (Fig.~\ref{fig:spectra}) that can provide
further constraints on the abundances, ionization state and dynamics
of the accretion disc. These features will also strongly depend on the
density of the disc because of the importance of collisional effects
and free-free emission/absorption. The spectra shown in
Fig.~\ref{fig:spectra} were calculated assuming a density of
$n_{\mathrm{H}}=10^{15}$~cm$^{-3}$, appropriate for the outer regions
of the disc. The density at a distance of a few gravitational radii
from the neutron star is expected to be $> 10^{21}$~cm$^{-3}$, even
for a radiation-pressure dominated disc \citep{ss73}, which is beyond
the range of validity for the reflection code. To illustrate the
effects of higher density on the soft X-ray spectrum,
Figure~\ref{fig:densitycompare} compares three reflection spectra
computed with $n_{\mathrm{H}}=10^{15}$~cm$^{-3}$ with ones computed at
$10^{18}$~cm$^{-3}$, where the code remains valid.  
The figure shows very little difference in the spectra in the
\textit{RXTE} band including the \fe\ line. This implies that the \fe\
EWs computed from the $n_{\mathrm{H}}=10^{15}$~cm$^{-3}$ models
(Fig.~\ref{fig:ews}) should still be useful guidelines for future
observations. On the other hand, the three order of magnitude increase
in density significantly alters the soft X-ray features predicted by
the reflection spectra. The differences mainly lie at energies $<
0.5$~\kev, where the continuum level is raised due to the increase in
free-free emission and absorption. As a result, the EWs of many of the
emission lines have been substantially decreased. The expectation is
that increasing the density further will continue to raise the
continuum level at soft energies, as more line emission is absorbed by
the continuum, and decrease the EW of the remaining emission
lines. Clearly, the soft X-ray emission features are a strong diagnostic
for the accretion disc density.  It would be therefore be very
interesting to obtain rapid, time-resolved broad-band spectra of X-ray
bursts to exploit the information in those spectral features. In the
near future only the XRT onboard the \textit{Swift} satellite will
have these features, plus the necessary rapid response
capability. Unfortunately, soft X-ray spectroscopy of burst sources
will be challenging. LMXBs are old systems and many of them lie toward
to the Galactic center, thus suffering from significant absorption due
to neutral hydrogen along the line of sight. However, there are a
number of LMXBs which reside in globular clusters that may provide
lower extinction columns and allow sensitive soft X-ray
observations. Rapid follow-up of X-ray bursts from LMXBs globular
clusters therefore provide the best chance of utilizing the full
information obtained in the reflection spectra.

\section*{Acknowledgments}
The author thanks R. Ross and A. Fabian for advice on the reflection
models, and acknowledges financial support from the Natural Sciences
and Engineering Research Council of Canada.


\bsp 

\label{lastpage}


\begin{thebibliography}{}
\bibitem[\protect\citeauthoryear{Angelini \etal}{1995}]{ang95}
  Angelini L., White N.E., Nagase F., Kallman T.R., Yoshida A.,
  Takeshima T., Becker C., Paerels F., 1995, ApJ, 449, L41
\bibitem[\protect\citeauthoryear{Asai \etal}{2000}]{adnm00} Asai K.,
  Dotani T., Nagase F., Mitsuda K., 2000, ApJS, 131, 571
\bibitem[\protect\citeauthoryear{Ballantyne \&
    Strohmayer}{2004}]{bs04} Ballantyne D.R., Strohmayer T.E., 2004,
    ApJ, 601, L105
\bibitem[\protect\citeauthoryear{Ballantyne, Iwasawa \&
Fabian}{Ballantyne \etal}{2001a}]{bif01} Ballantyne D.R., 
Iwasawa K., Fabian, A.C., 2001a, MNRAS, 323, 506
\bibitem[\protect\citeauthoryear{Ballantyne, Ross \&
Fabian}{Ballantyne \etal}{2001b}]{brf01} Ballantyne D.R., Ross R.R.,
Fabian A.C., 2001b, MNRAS, 327, 10
\bibitem[\protect\citeauthoryear{Ballantyne, Vaughan \&
    Fabian}{Ballantyne \etal}{2003}]{bvf03} Ballantyne D.R., Vaughan
    S., Fabian A.C., 2003, MNRAS, 342, 239
\bibitem[\protect\citeauthoryear{Ballantyne, Turner \&
    Blaes}{Ballantyne \etal}{2004}]{btb04} Ballantyne D.R., Turner
    N.J., Blaes O.M., 2004, ApJ, in press (astro-ph/0311390)
\bibitem[\protect\citeauthoryear{Barrio, Done \& Nayakshin}{Barrio
    \etal}{2003}]{bdn03} Barrio F.E., Done C., Nayakshin S.,
    2003, MNRAS, 342, 557
\bibitem[\protect\citeauthoryear{Cornelisse \etal}{2000}]{corn00}
  Cornelisse R., Heise J., Kuulkers E., Verbunt F., in 't Zand J.J.M.,
  2000, A\&A, 357, L21
\bibitem[\protect\citeauthoryear{Cottam \etal}{2001}]{cott01} Cottam
  J., Sako M., Kahn S.M., Paerels F., Liedahl D.A., 2001, ApJ, 557,
  L101
\bibitem[\protect\citeauthoryear{Cottam, Paerels \&
    Mendez}{2002}]{cpm02} Cottam J., Paerels F., Mendez M., 2002,
    Nature, 420, 51
\bibitem[\protect\citeauthoryear{Day \& Done}{1991}]{dd91} Day C.S.R.,
  Done C., 1991, MNRAS, 253, 35\textsc{p}
\bibitem[\protect\citeauthoryear{Day, Fabian \& Ross}{1992}]{dfr92} Day
  C.S.R., Fabian A.C., Ross R.R., 1992, MNRAS, 257, 471
\bibitem[\protect\citeauthoryear{De Rosa \etal}{2002}]{der02} De Rosa
  A., Piro L., Fiore F., Grandi P., Maraschi, L., Matt G.,
  Nicastro F., Petrucci P.O., 2002, A\&A, 387, 838 
\bibitem[\protect\citeauthoryear{Di Matteo}{1998}]{dm98} Di Matteo T.,
  1998, MNRAS, 299, L15 
\bibitem[\protect\citeauthoryear{Fabian \etal}{1989}]{fab89}Fabian
  A.C., Rees M.J., Stella L., White N.E., 1989, MNRAS, 238, 729
\bibitem[\protect\citeauthoryear{Foster, Ross \& Fabian}{1987}]{frf87}
  Foster A.J., Ross R.R., Fabian A.C., 1987, MNRAS, 228, 259
\bibitem[\protect\citeauthoryear{Galeev, Rosner \& Vaiana}{Galeev
\etal}{1979}]{gal79} Galeev A.A., Rosner R., Vaiana G.S., 1979, ApJ,
229, 318 
\bibitem[\protect\citeauthoryear{George \& Fabian}{1991}]{gf91} George
I.M., Fabian A.C., 1991, MNRAS, 249, 352
\bibitem[\protect\citeauthoryear{Gierli\'{n}ski \etal}{1999}]{gie99}
  Gierli\'{n}ski M., Zdziarski A.A., Poutanen J., Coppi P.S., Ebisawa
  K., Johnson W.N., 1999, MNRAS, 309, 496
\bibitem[\protect\citeauthoryear{Haardt \&
Maraschi}{1993}]{haa93}Haardt F., Maraschi L., 1993, ApJ, 413, 507
\bibitem[\protect\citeauthoryear{Haardt, Maraschi \&
Ghisellini}{Haardt \etal}{1994}]{haa94} Haardt F., Maraschi L.,
Ghisellini G., 1994, ApJ, 432, L95
\bibitem[\protect\citeauthoryear{Hoffman, Lewin \& Doty}{Hoffman
    \etal}{1977}]{hld77} Hoffman J.A., Lewin W.H.G., Doty J., 1977,
    ApJ, 217, L23
\bibitem[\protect\citeauthoryear{Jimenez-Garate, Raymond \&
    Liedahl}{Jimenez-Garate \etal}{2002}]{jrl02} Jimenez-Garate M.A.,
    Raymond J.C., Liedahl D.A., 2002, ApJ, 581, 1297
\bibitem[\protect\citeauthoryear{Kallman \etal}{2003}]{kabc03} Kallman
  T.R., Angelini L., Boroson B., Cottam J., 2003, ApJ, 583, 861
\bibitem[\protect\citeauthoryear{Ko \& Kallman}{1994}]{kk94} Ko Y.-K.,
  Kallman T.R., 1994, ApJ, 431, 273
\bibitem[\protect\citeauthoryear{Kuulkers}{2003}]{kuul03}Kuulkers
  E., 2003, in The Restless High-Energy Universe, eds. van den Heuvel
  E.P.J., in 't Zand J.J.M., Wijers R.A.M.J., in press
  (astro-ph/0310402)
\bibitem[\protect\citeauthoryear{Liu, van Paradijs \& van den
    Heuvel}{Liu \etal}{2001}]{lph01} Liu Q.Z., van Paradijs J., van
    den Heuvel E.P.J., 2001, A\&A, 368, 1021
\bibitem[\protect\citeauthoryear{Longinotti \etal}{2003}]{lon03}
  Longinotti A.L., Cappi M., Nandra K., Dadina M., Pellegrini
  S., 2003, A\&A, 410, 471
\bibitem[\protect\citeauthoryear{Magnier \etal}{1989}]{mag89} Magnier
  E., Lewin W.H.G., van Paradijs J., Tan J., Penninx W., Damen E.,
  1989, MNRAS, 237, 729
\bibitem[\protect\citeauthoryear{Matt, Perola \& Piro}{Matt
\etal}{1991}]{mpp91} Matt G., Perola G.C., Piro L., 1991, A\&A, 247,
25
\bibitem[\protect\citeauthoryear{Meyer \&
    Meyer-Hofmeister}{1982}]{mmh82} Meyer F., Meyer-Hofmeister E.,
    1982, A\&A, 106, 34
\bibitem[\protect\citeauthoryear{Miller \etal}{2004}]{mil03} Miller
  J.M., \etal, 2004, ApJ, in press (astro-ph/0307394) 
\bibitem[\protect\citeauthoryear{Morrison \& McCammon}{1983}]{mcm83}
Morrison R., McCammon D., 1983, ApJ, 270, 119
\bibitem[\protect\citeauthoryear{Nakamura, Inoue \& Tanaka}{Nakamura
    \etal}{1988}]{nit88} Nakamura N., Inoue H., Tanaka Y., 1988, PASJ,
    40, 209
\bibitem[\protect\citeauthoryear{Nandra \& Pounds}{1994}]{np94} Nandra
K., Pounds K.A., 1994, MNRAS, 268, 405 
\bibitem[\protect\citeauthoryear{Nayakshin \& Kallman}{2001}]{nk01}
Nayakshin S., Kallman T.R., 2001, ApJ, 546, 406
\bibitem[\protect\citeauthoryear{Nayakshin, Kazanas \&
Kallman}{Nayakshin \etal}{2000}]{nkk00} Nayakshin S., Kazanas D.,
Kallman T.R., 2000, ApJ, 537, 833
\bibitem[\protect\citeauthoryear{Orr \etal}{2001}]{orr01} Orr A., Barr
P., Guainazzi M., Parmar A.N., Young, A.J., 2001, A\&A, 376, 413
\bibitem[\protect\citeauthoryear{Pounds \etal}{1990}]{pou90} Pounds
K.A., Nandra K., Stewart G.C., George I.M., Fabian A.C., 1990, Nature,
344, 132
\bibitem[\protect\citeauthoryear{Ross \& Fabian}{1993}]{ros93} Ross
R.R., Fabian A.C., 1993, MNRAS, 261, 74
\bibitem[\protect\citeauthoryear{Ross, Weaver \& McCray}{Ross et
    al}{1978}]{rwm78} Ross R.R., Weaver R., McCray R., 1978, ApJ, 219, 292
\bibitem[\protect\citeauthoryear{Ross, Fabian \& Young}{Ross
\etal}{1999}]{ros99} Ross R.R., Fabian A.C., Young A.J., 1999, MNRAS,
306, 461
\bibitem[\protect\citeauthoryear{R\'{o}\.{z}a\'{n}ska \etal}{2002}]{roz02}
R\'{o}\.{z}a\'{n}ska A., Dumont A.-M., Czerny B., Collin S., 2002,
MNRAS, 332, 799
\bibitem[\protect\citeauthoryear{Rybicki \& Lightman}{1979}]{ryl79}
Rybicki G.B., Lightman A.P., 1979, Radiative Processes in
Astrophysics, Wiley
\bibitem[\protect\citeauthoryear{Schulz}{1999}]{sch99} Schulz N.S.,
  1999, ApJ, 511, 304
\bibitem[\protect\citeauthoryear{Schulz \etal}{2001}]{sch01} Schulz
  N.S., Chakrabarty D., Marshall H.L., Canizares C.R., Lee J.C., Houck
  J., 2001, ApJ, 563, 941
\bibitem[\protect\citeauthoryear{Shakura \& Sunyaev}{1973}]{ss73}
  Shakura N.I., Sunyaev R.A., 1973, A\&A, 24, 337
\bibitem[\protect\citeauthoryear{Shapiro, Lightman \& Eardley}{Shapiro
\etal}{1976}]{sle76} Shapiro S.L., Lightman A.P., Eardley D.M., 1976,
ApJ, 204, 187
\bibitem[\protect\citeauthoryear{Smale \etal}{1993}]{sma93} Smale
  A.P., \etal, 1993, ApJ, 410, 796
\bibitem[\protect\citeauthoryear{Strohmayer \& Brown}{2002}]{sb02}
  Strohmayer T.E., Brown E.F., 2002, ApJ, 566, 1045
\bibitem[\protect\citeauthoryear{Strohmayer \& Bildsten}{2003}]{sb03}
  Strohmayer T.E., Bildsten L., 2003, in Compact Stellar X-ray
  Sources, eds. Lewin W.H.G., van der Klis M. (Cambridge: Cambridge
  University Press), in press (astro-ph/0301544)
\bibitem[\protect\citeauthoryear{Swank \etal}{1977}]{swa77} Swank
  J.H., Becker R.H., Boldt E.A., Holt S.S., Pravdo S.H., Serlemitsos
  P.J., 1977, ApJ, 212, L73
\bibitem[\protect\citeauthoryear{Waki \etal}{1984}]{wak84} Waki I.,
  \etal, 1984, PASJ, 36, 819
\bibitem[\protect\citeauthoryear{Zdziarski, Lubi\'{n}ski \&
Smith}{Zdziarski \etal}{1999}]{zdz99} Zdziarski A.A.,
Lubi\'{n}ski P., Smith D.A., 1999, MNRAS, 303, 11
\bibitem[\protect\citeauthoryear{Zdziarski \etal}{2003}]{zlgr03}
  Zdziarski A.A., Lubi\'{n}ski P., Gilfanov M., Revnivtsev M., 2003,
  MNRAS, 342, 355
\bibitem[\protect\citeauthoryear{\.{Z}ycki \etal}{1994}]{zyc94}
\.{Z}ycki P.T., Krolik J.H., Zdziarski A.A., Kallman T.R., 1994, ApJ, 437, 597
\end{thebibliography}
\end{document}